\documentclass[aip,jcp]{revtex4}

\usepackage{amsmath,amssymb,graphicx}
\usepackage{epsfig}
\usepackage{graphicx}
\usepackage{enumerate}
\usepackage{graphicx}
\usepackage{color}
\usepackage{amsmath}
\usepackage{comment}
\usepackage{placeins}
\usepackage{dcolumn}
\usepackage{bm}
\usepackage{hyperref}
\usepackage{epsf}
\usepackage{subfigure}
\usepackage{epstopdf}%
\usepackage{amsfonts}

\newcommand{\bea}{\begin{eqnarray}}
\newcommand{\eea}{\end{eqnarray}}

\def\gg{{\rangle\rangle}}
\def\ll{{\langle\langle}}

\usepackage{color}

\begin{document}

\title{Thermoelectricity in molecular junctions with harmonic and anharmonic modes}

\author{Bijay Kumar Agarwalla$^1$, Jian-Hua Jiang $^2$, Dvira Segal$^1$}
\affiliation{$^1$Chemical Physics Theory Group, Department of Chemistry, and Centre for Quantum Information and Quantum Control,
University of Toronto, 80 Saint George St. Toronto, Ontario, Canada M5S 3H6}
\affiliation{$^2$Department of Physics, Soochow University, 1 Shizi Street, Suzhou 215006, China}



\begin{abstract}
We study charge and energy transfer in two-site molecular electronic junctions
in which electron transport is assisted by a vibrational mode.
To understand the role of mode harmonicity/anharmonicity on transport behavior,
we consider two limiting situations: (i) the mode is assumed harmonic,
(ii) we truncate the mode spectrum to include only two levels,
to represent an anharmonic molecular mode.
Based on the models' cumulant generating functions,
we analyze the linear-response and nonlinear performance of these junctions  and demonstrate
that while the electrical and thermal conductances are sensitive to whether the mode is harmonic/anharmonic,
the Seebeck coefficient, the thermoelectric figure-of-merit, and the thermoelectric efficiency beyond linear response,
 conceal this information.
\end{abstract}
\maketitle


\section{Introduction}

Molecular electronic junctions offer a rich playground for exploring basic and practical questions in
quantum transport, such as the interplay between electronic and nuclear dynamics in nonequilibrium situations.
Theoretical and computational efforts based on minimal model Hamiltonians are largely focused on
the Anderson impurity dot model which consists a single molecular electronic orbital directly
coupled to biased metal leads, as well as to a particular vibrational mode  \cite{nitzan}.
Since the same molecular orbital is assumed to extend both contacts,
the model allows simulations of transport characteristics in conjugated molecular junctions with delocalized electrons.

\begin{figure}
\caption{
Scheme of the D-A model considered in this work.
A two-site (D, A) electronic junction is
coupled to either (a) a harmonic molecular mode, or (b) an anharmonic
mode, represented by a two-state system.
The molecular mode may further relax its energy
to a phononic thermal reservoir (not depicted here), maintained at temperature $T_{ph}$.
}
\label{scheme}
\includegraphics[width=12cm]{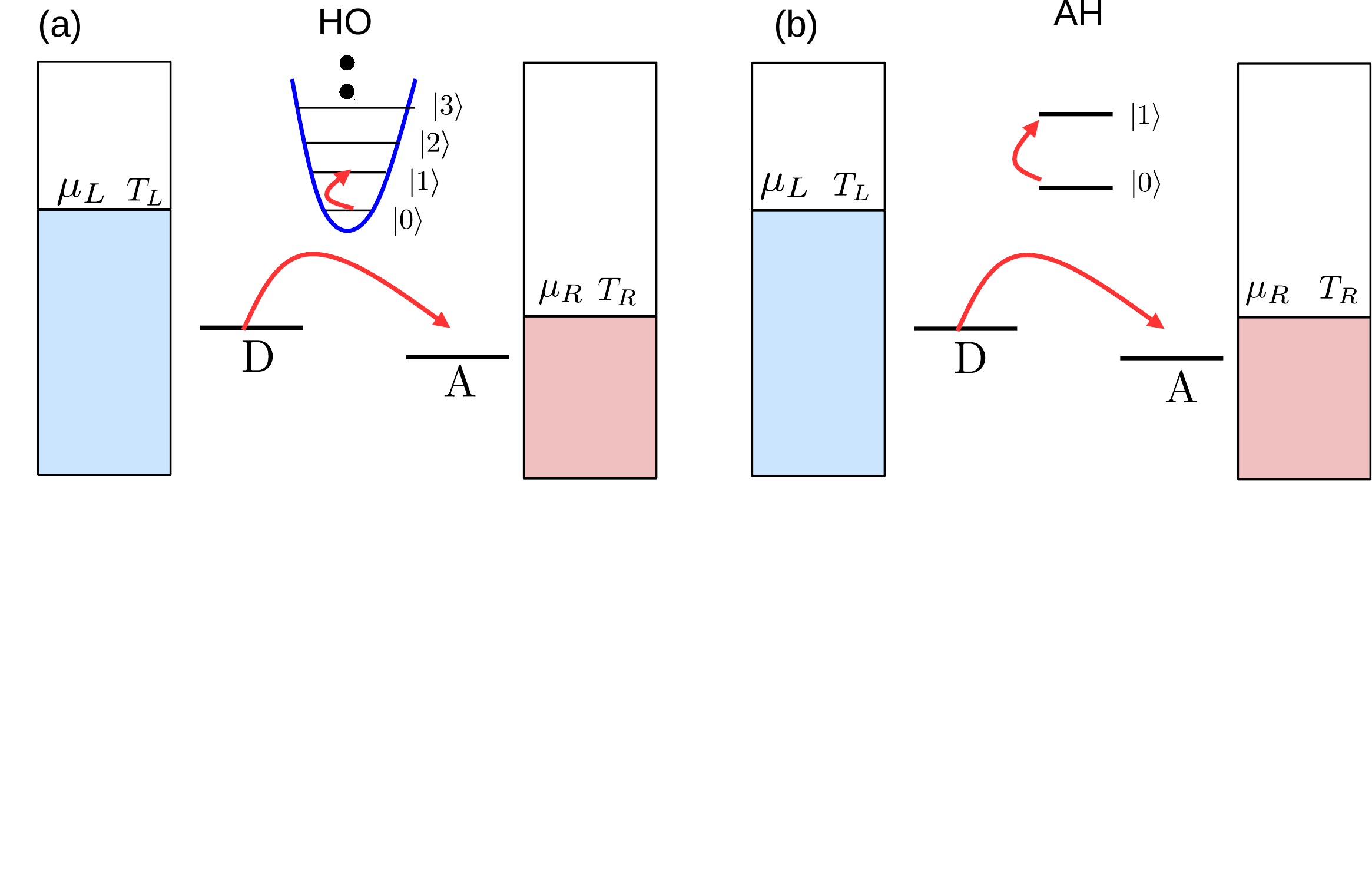}
\vspace{-27mm}
\end{figure}

In this work, we focus on a different class of  molecular junctions as depicted in Fig. \ref{scheme}.
In such systems, two electronic levels are coupled via a weak tunneling element,
but electrons may effectively hop between these electronic states when interacting with
a vibrational mode. This could  e.g.,  correspond to a torsional motion bringing orthogonal $\pi$ systems
into an overlap as in the  2,2'- dimethylbiphenyl (DMBP) molecule
recently examined in Refs. \cite{thossE,Markussen,SimineJPC15}.
This model is related to the original
Aviram-Ratner construction for a donor-acceptor molecular rectifier \cite{Aviram},
thus we identify the two states here as D and A, see Fig. \ref{scheme}, and refer to the model as the ``D-A junction".
More recently, this construction  was employed for exploring vibrational heating and
instability under a large bias voltage \cite{Lu,SegalSF1,SegalSF2}.
The system is also referred to as the ``dimer molecular junction" \cite{Nori}, or an ``open spin-boson model" \cite{Brandes}
(where the spin here represents the D and A states, the bosons correspond to the molecular vibrational modes,
and the system is open- coupled to metal leads).
It was utilized to study charge transfer in donor-bridge-acceptor organic molecules  \cite{Berlin} and
organic molecular semiconductors \cite{noncondon}, as well as thermoelectric effects
in quantum dot devices \cite{JianHua1,JianHua2}. 
Recently,  Erpenbeck {\it et al.} had provided a thorough
computational study of transport characteristics
with nondiagonal (or nonlocal) as well as diagonal (local) electron-vibration interactions \cite{thoss15}.
Here, in contrast, we simplify
the junction model and omit the contribution of direct tunneling between the D and A units.
This simplification allows us to derive a closed (perturbative) expression
for the cumulant generating function (CGF) of the model, which contains comprehensive information over
transport characteristics. 


Measurements of charge current and electrical conductance in single molecules hand over detailed
energetic and dynamical information \cite{latha}.
Complementing electrical conductance measurements, the thermopower, a linear response quantity,  also referred to as the Seebeck
coefficient, is utilized as an independent tool for probing the energetics of molecular junctions, see for example Refs.
\cite{reddy,Reddy9,Reddy10,Reddy11,Tao,MalenE08, Latha13,reddy14}.
Experimental efforts identified orbital hybridization, contact-molecule energy coupling and geometry, and whether
the conductance is HOMO or LUMO dominated.
More generally, the thermoelectric performance beyond linear response is of interest, with
the two metal leads maintained at (largely) different temperatures and chemical potentials.

What information can linear and nonlinear thermoelectric transport coefficients reveal on molecular junctions?
Specifically, can they expose the underlying electron-phonon interactions and the
characteristics of the vibrational modes participating in the process?
Focusing on the challenge of efficient thermoelectric systems,
how should we tune molecular parameters to improve heat to work conversion efficiency?
These questions were examined in recent studies, a non-exhaustive list includes
\cite{Koch,SegalE05,Kamil,ChenVib,polaron,Cata,Li12,Baowen,Lili,Ora,Ora2,AIP,JianHua1,JianHua2}.

We focus here on the effect of vibrational anharmonicity on thermoelectric transport within the D-A model.
To explore this issue,
two limiting variants of the basic  construction are examined, as displayed in  Fig. \ref{scheme}:
(a) The vibration is harmonic in the so-called ``harmonic oscillator" (HO) model.
(b) To learn about deviations from the harmonic picture,   
we truncate the vibrational spectrum to include only its two lowest levels,
 constructing the ``anharmonic" (AH) mode model.
In a different context, the AH model could represent transport through molecular magnets,
in which electron transfer is controlled by a spin impurity \cite{spin2}.

The complete information over transport behavior
is contained in the respective cumulant generating functions,
which we provide here for the HO and the AH models, valid under the approximation of weak electron-vibration interactions.
From the CGFs, we derive expressions for charge and heat currents,
and study the linear and nonlinear thermoelectric performance of the junctions.
Focusing on the role of vibrational anharmonicity, we find that while it significantly influences the electrical and
thermal conductances, nevertheless in the present model it does not affect heat-to-work conversion efficiency.




\section{Model}

We consider a two-site junction, where
electron hopping between the D and A
electronic states (creation operators $c_d^{\dagger}$ and $c_a^{\dagger}$, respectively)
is assisted by a vibrational mode.
The total Hamiltonian is  written as
\bea
H&=&
\epsilon_d c_d^{\dagger} c_d + \epsilon_a c_a^{\dagger} c_a + H_{vib} + H_{el-vib}
\nonumber\\
&+& \sum_{l\in L} {\epsilon}_l c_l^{\dagger} c_l + \sum_{r\in R} {\epsilon}_r c_r^{\dagger} c_r.
\nonumber \\
&+& \sum_{l \in L} v_{l} (c_l^{\dagger} c_d + c_d^{\dagger} c_l) + \sum_{r \in R} v_{r} (c_r^{\dagger} c_a + c_{a}^{\dagger} c_r).
\label{eq:H}
\eea
The molecular electronic states of energies $\epsilon_{d,a}$ are hybridized with their adjacent metals,
collection of noninteracting electrons, by hopping elements $v_{l}$ and $v_r$.
Here $c_{j}^{\dagger}$ ($c_j$) is a fermionic creation (annihilation) operator.
The electronic Hamiltonian [Eq. (\ref{eq:H}), excluding $H_{vib}+H_{el-vib}$]
can be diagonalized and expressed in terms of new fermionic operators $a_l$ and $a_r$,
%
{\bea
H_{el} = \sum_{l} \epsilon_l a_l^{\dagger} a_l + \sum_{r} \epsilon_r a_r^{\dagger} a_r.
\label{eq:Hel}
\eea
%
The molecular operators can be expanded in the new basis as,
$c_d = \sum_{l} \gamma_l a_l, \quad c_a = \sum_{r} \gamma_r a_r$.
The $\gamma_{l,r}$ coefficients satisfy
$\gamma_l= v_l\Big/\left(\epsilon_l-\epsilon_d - \sum_{l'} \frac{v_{l'}^2}{\epsilon_l-\epsilon_{l'} + i \delta}\right)$
\cite{SegalSF1}.
The operators $c_l$ and $c_r$ can be expressed in terms of the new basis
as $c_l =\sum_{l'} \eta_{l l'} a_{l'}$, 
with
$\eta_{ll'} = \delta_{ll'}- \frac{v_l \gamma_{l'}}{\epsilon_l-\epsilon_{l'} + i \delta}$.
Similar expressions hold for the $r$ set.

Back to Eq. (\ref{eq:H}), $H_{vib}$ and $H_{el-vib}$ represent the Hamiltonians of the molecular vibrational mode and its
coupling to electrons, respectively.  We assume an ``off-diagonal" interaction with
electron hopping between local sites assisted by the vibrational mode.
Assuming a local harmonic mode we write
\bea
H_{vib}&=& \omega_{0} b_{0}^{\dagger} b_0, \nonumber \\
H_{el-vib} &=& g [c_{d}^{\dagger} c_{a} + c_{a}^{\dagger} c_d ]  (b_0^{\dagger} + b_0),
\label{eq:vib}
\eea
with $b_0$ ($b_0^{\dagger})$ as the annihilation (creation) operator for the vibrational mode of
frequency $\omega_0$, $g$ is the coupling parameter.
The Hamiltonian (\ref{eq:H}) then becomes
\bea
H_{HO}= H_{el} + \omega_0 b_0^{\dagger} b_0 + g \sum_{l\in L,r\in R} \big[\gamma_l^* \gamma_r a_l^{\dagger} a_r + h.c.\big](b_0^{\dagger} + b_0).
\label{eq:HO}
\eea
The second model considered here includes an anharmonic two-state mode.
It is convenient to represent it with the Pauli matrices $\sigma_{x,y,z}$,
and write the total Hamiltonian for the junction as
\bea
H_{AH}= H_{el} + \frac{\omega_0}{2} \sigma_z + g \sum_{l\in L,r\in R} \big[\gamma_l^* \gamma_r a_l^{\dagger} a_r + h.c.\big]\sigma_x.
\label{eq:AH}
\eea
%
The two models, Eqs. (\ref{eq:HO}) and (\ref{eq:AH}), describe electron-hole pair generation/annihilation
by de-excitation/excitation of an ``impurity" (vibrational mode).
The left and right reservoirs defining $H_{el}$ in Eq. (\ref{eq:Hel}) are characterized by a structured density of states
since we had absorbed the D state in the $L$ terminal, and similarly, the A level in the $R$ metal.
These electronic reservoirs are prepared in a thermodynamic state of
temperature $T_{\nu}=1/(k_B\beta_{\nu})$ and chemical potential $\mu_{\nu}$, $\nu=L,R$, set relative to the equilibrium chemical potential
$\mu_F=0$.
In our description we work with $\hbar =1$ and $e=1$. Units are revived in simulations.

\section{Transport}

The complete information over steady state charge and energy transport properties of molecular junctions
is delivered by the so-called
cumulant generating function $\cal G$, defined in terms of the characteristic function ${\cal Z}$ as
${\cal G}(\lambda_e, \lambda_p)= \lim_{t \to \infty} \frac{1}{t} \ln {\cal Z}(\lambda_e, \lambda_p)$
with ${\cal Z}(\lambda_e, \lambda_p) = \langle e^{i \lambda_e H_R + i \lambda_p N_R} \, e^{-i \lambda_e H_R^H(t) - i \lambda_p N_R^H(t)} \rangle$.
Here $\lambda_e, \lambda_p$ are counting fields for energy and particles, respectively,
defined for the right lead measurement,
$t$ is the final measurement time.
The operators in this definition are
 $N_R= \sum_r a_r^{\dagger} a_r$,  the number operator for the total charge in the right lead,
 and  similarly $H_R= \sum_r \epsilon_r a_r^{\dagger} a_r$ as the total energy in the same compartment.
The superscript $H$ identifies the Heisenberg picture, with operators evolving with respect to
the full Hamiltonian.
While closed results for the CGF can be derived for junctions of noninteracting particles \cite{FCS},
it is challenging to calculate this function analytically for models with interactions,
see for example Ref. \cite{ORA}.
Our simplified D-A model is one of the very few many-body models which can be solved analytically.

The CGF of the harmonic-mode junction (\ref{eq:HO}) can be derived using
the nonequilibrium Green's function (NEGF) technique  \cite{NEGFbijay1,NEGFbijay2}
 assuming weak interaction between electrons and the particular
vibration, employing the random phase approximation (RPA) \cite{RPA1,RPA2}.
This scheme involves a summation
over a particular set (infinite) of diagrams (ring type) in the perturbative series,
taking into account all electron scattering processes which are facilitated by
the absorption or emission of a {\it single} quantum $\omega_0$.
Physically, this summation collects
not only sequential tunnelling electrons, but all coordinated multi-tunnelling processes,
albeit with each electron interacting with the mode to the lowest order. 
The derivation of the CGF is nontrivial,
and it is included in a separate communication  \cite{BijayCGF}. Here we
provide the final result
\bea
{\cal G}_{HO}(\lambda_e,\lambda_p) =
\frac{1}{2}(k_{d} - k_{u}) - \frac{1}{2}\sqrt{(k_{u} + k_{d})^2 - 4 k_{u}^{\lambda} k_{d}^{\lambda}}.
\label{eq:CGFHO}
\eea
The CGF of the AH  model (\ref{eq:AH}) can be derived based on a counting-field dependent master equation approach
\cite{SegalSF1,BijayCGF},
\bea
{\cal G}_{AH}(\lambda_e,\lambda_p) =
-\frac{1}{2}(k_u + k_d) + \frac{1}{2}\sqrt{(k_{u} - k_{d})^2 + 4\, k_{u}^{\lambda} k_{d}^{\lambda}}.
\label{eq:CGFAH}
\eea
Both expressions  are correct to second-order in the electron-vibration coupling $g$.
It is remarkable to note on the similarity of these expressions
which were derived from separate approaches.
We use the short notation $\lambda=(\lambda_p,\lambda_e)$, where
the counting fields are defined for right-lead measurements.
It can be proved that our CGFs satisfy the fluctuation symmetry \cite{fluctR}
\bea
{\cal G}(\lambda_e,\lambda_p)= {\cal G} (-\lambda_e-i\Delta\beta, -\lambda_p-i(\beta_L\mu_L-\beta_R\mu_R) ),
\label{eq:FT}
\eea
with $\Delta \beta=\beta_R-\beta_L$.
This result is not  trivial: Schemes involving truncation of interaction elements may
leave out terms inconsistently with the fluctuation symmetry.
%
%
Equations (\ref{eq:CGFHO}) and (\ref{eq:CGFAH}) are expressed in terms of an upward (excitation) $k_{u}^{\lambda}$
and a downward (de-excitation) $k_{d}^{\lambda}$ rates between vibrational states. The rates are additive in
the two baths,
\bea
k_{d}^{\lambda}= [k_{d}^{\lambda}]^{L \to R} + \big[k_{d}^{\lambda}\big]^{R \to L},
\label{eq:r0}
\eea
and  obey the relation $k_{u}^{\lambda}=  k_{d}^{\lambda} [\omega_0 \to -\omega_0]$.
They are given by \cite{SegalSF1,BijayCGF}
%
\bea
[k_{d}^{\lambda}]^{L \to R} &=& \int_{-\infty}^{\infty} \frac{d\epsilon}{2\pi} f_L(\epsilon) (1-f_R(\epsilon+\omega_0)) J_L(\epsilon) J_R(\epsilon+\omega_0) \, e^{-i \lambda_p-i(\epsilon + \omega_0) \lambda_e}, \nonumber \\
\big[k_{d}^{\lambda}\big]^{R \to L} &=&  \int_{-\infty}^{\infty} \frac{d\epsilon}{2\pi} f_R(\epsilon) (1-f_L(\epsilon+\omega_0)) J_R(\epsilon) J_L(\epsilon+\omega_0) \, e^{i \lambda_p+i\epsilon \lambda_e}.
\label{eq:rates}
\eea
%
The rates $k_d$ and $k_u$ are evaluated from these expressions at $\lambda=0$;
 $f_{\nu}(\epsilon)=[\exp(\beta_{\nu}(\epsilon-\mu_{\nu}))+1]^{-1}$ is
the Fermi-Dirac distribution function of the $\nu=L,R$ lead.
The properties of the molecular junction are embedded within
the spectral density functions, peaked around the molecular electronic energies $\epsilon_{d,a}$
with the broadening $\Gamma_{\nu}(\epsilon)=2\pi \sum_{k\in \nu} v_k^2\delta(\epsilon-\epsilon_k)$,
\bea
J_L(\epsilon) &=& g \frac{\Gamma_L(\epsilon)}{(\epsilon-\epsilon_d)^2 + (\Gamma_L(\epsilon)/2)^2},
\nonumber\\
J_R(\epsilon) &=& g \frac{\Gamma_R(\epsilon)}{(\epsilon-\epsilon_{a})^2 + (\Gamma_R(\epsilon)/2)^2}.
\label{eq:JJ}
\eea
%
These expressions are reached through the diagonalization procedure of the electronic Hamiltonian
while ignoring the real principal value term, responsible for a small energy shift of $\epsilon_{d,a}$ \cite{SegalSF1}.
In what follows we take
$\Gamma_{\nu}$ as a constant independent of energy and assume broad bands with a large cutoff $\pm D$, the largest
energy scale in the problem.

We obtain currents and high order cumulants
by taking derivatives of the CGF with respect to the counting fields.
The particle $\langle I_p\rangle$ and energy $\langle I_e\rangle $ current are given by
\bea
&&\langle I_p\rangle \equiv \frac{\langle N \rangle}{t} =
\frac{\partial {\cal G}(\lambda_e,\lambda_p)}{\partial (i \lambda_p)}\Big{|}_{\lambda_e=\lambda_p=0}, \quad
\nonumber\\
&&
\langle I_e\rangle \equiv \frac{\langle Q \rangle}{t} = \frac{\partial {\cal G}(\lambda_e,\lambda_p)}{\partial (i \lambda_e)}\Big{|}_{\lambda_e=\lambda_p=0}.
\label{eq:current}
\eea
%
%
%
%
After some manipulations  we reach
the compact form for the harmonic ($-$) and anharmonic ($+$) models,
\bea
&&\langle I_p^{AH/HO} \rangle =
2 \, \frac{k_{d} ^{R \to L} k_{ u}^{R \to L} - k_{d}^{L \to R} k_{u}^{L \to R}}
{k_{d} \pm k_{u}}, \nonumber \\
&&\langle I_e^{AH/HO} \rangle =
\frac{k_{d}\,\big[ \partial_{(i \lambda_e)} k_{u}^{\lambda}|_{\lambda_e=0}\big]+ k_{u} \big[ \partial_{(i \lambda_e)} k_{d}^{\lambda}|_{\lambda_e=0}\big]}
{k_{d} \pm k_{u}}.
\label{eq:I}
\eea
The rates are given by Eq. (\ref{eq:rates}) with $\lambda=0$.
It is notable that the only difference between the HO and AH models is the sign in the denominator.
Note that we did not simplify the expression for the energy current $\langle I_e^{AH/HO} \rangle$ above;
the derivatives return energy transfer rates analogous
to Eq. (\ref{eq:rates}), only with an additional energy variable in the integrand.
While figures below only display quantities related to charge and energy currents,
it is useful to emphasize that the CGF contains
information on fluctuations of these currents.
For example, the zero-frequency noise for charge current is given from
\bea
&&\langle S_p \rangle \equiv \frac{\langle \langle N^2 \rangle \rangle}{t} =
\frac{\partial^2 {\cal G}(\lambda_e,\lambda_p)}{\partial (i \lambda_p)^2}\Big{|}_{\lambda_e=\lambda_p=0},
%
\eea
where  $\langle \langle N^2 \rangle \rangle = \langle N^2\rangle - \langle N \rangle^2$
is the second cumulant. 
%

Our derivation is based on the diagonal representation of the electronic Hamiltonian,
thus the occupations of the molecular electronic states D and A follow the
Fermi function by construction.
In the weak coupling limit employed here, the back-action of the vibrational degrees of freedom on the electronic
distribution is not included.
While in other models \cite{Park} this back-action may be significant, here we argue that its role is rather small:
Recent numerically-exact path integral simulations \cite{SegalSF2} testify that this type of
quantum master equation very well performs at
weak-intermediate electron-vibration coupling, justifying our scheme.
Note that in path integral simulations \cite{SegalSF2}
the states D and A were absorbed into the metal leads as well,
yet the electronic distribution was allowed to evolve in time,
naturally incorporating the back-effect of vibrations on the electronic distribution in the steady-state limit.


\section{Results}

We are interested in identifying signatures of mode harmonicity in transport characteristics.
We set the right contact as hot, $T_R>T_L$,
and write the electronic heat current extracted from the hot bath by
$\langle I_h\rangle = \langle I_e \rangle -\mu_R \langle I_p\rangle$.
The bias is applied such that $\mu_L>\mu_R$, thus
the macroscopic efficiency of a thermoelectric device, converting heat to work, is given by
\bea
\eta\equiv\frac{-\langle W\rangle }{\langle I_h\rangle}  = \frac{(\mu_L-\mu_R)\langle I_p \rangle }{\langle I_h \rangle }.
\label{eq:eta}
\eea
The device is operating as a thermoelectric engine when both charge and energy current flow from the hot (right) bath to the cold one.
Note that according to our conventions the currents are positive when flowing from the right contact to the left.
%

\subsection{Linear response coefficients}

In linear response, i.e., close to equilibrium, the charge current $\langle I_p\rangle $ and heat current $\langle I_h\rangle$
as obtained from  Eq. (\ref{eq:I}) can be
expanded to lowest order in the bias voltage $\Delta \mu =\mu_R-\mu_L=e V$
and temperature difference $\Delta T=T_R-T_L$. To re-introduce physical dimensions, we
multiply the charge current by  $e/\hbar$ and the heat current by $1/\hbar$.
The resulting expansions are cumbersome thus
we write them formally in terms of the coefficients $a_{i,j}$, ($i,j=h,p$),
%
\bea
\langle I_p\rangle =\frac{e}{\hbar}a_{p,p}\Delta \mu +\frac{e}{\hbar}a_{p,h}k_B\Delta T,
\nonumber\\
\langle I_h\rangle =\frac{1}{\hbar}a_{h,p}\Delta \mu +\frac{1}{\hbar}a_{h,h}k_B\Delta T.
\eea
Recalling that $\langle I_p\rangle=GV +GS\Delta T $ and
$\langle I_h\rangle=G\Pi V +(\Sigma +GS\Pi)\Delta T$
with $\Pi$ as Peltier coefficient \cite{Onsager31,casati13},
we identify the electrical conductance
$G=\frac{e^2}{\hbar}a_{p,p}$, the thermopower $S=\frac{k_B}{e}\frac{a_{p,h}}{a_{p,p}}$,
the electron contribution to the thermal conductance $\Sigma=\frac{k_B}{\hbar} \left(  a_{h,h}-\frac{a_{p,h} a_{h,p}}{a_{p,p}}\right)$
and the (dimensionless) thermoelectric figure of merit $ZT=\frac{GS^2}{\Sigma}T$ which determine the linear response thermoelectric
efficiency.
We obtain these coefficients numerically, by simulating Eq. (\ref{eq:I}) under small biases.


\begin{figure}
\caption{Linear response behavior of the D-A junction as a function of molecule-metal hybridization
with a harmonic mode (full) and
an anharmonic two-state system mode (dashed).
(a) Normalized electrical conductance $G/G_0$ with $G_0=e^2/h$, the quantum of conductance per channel per spin.
(b) Seebeck coefficient $S$.
(c) Electronic thermal conductance $\Sigma$, and (d) the figure of merit $ZT$.
Parameters are $\epsilon_0=0.01$, $\omega_0=0.02$, $g=0.01$ in eV, room temperature
$T=300$ K. We assumed flat bands with a constant density of states.
}
\label{fig2}
\includegraphics[width=7.5cm]{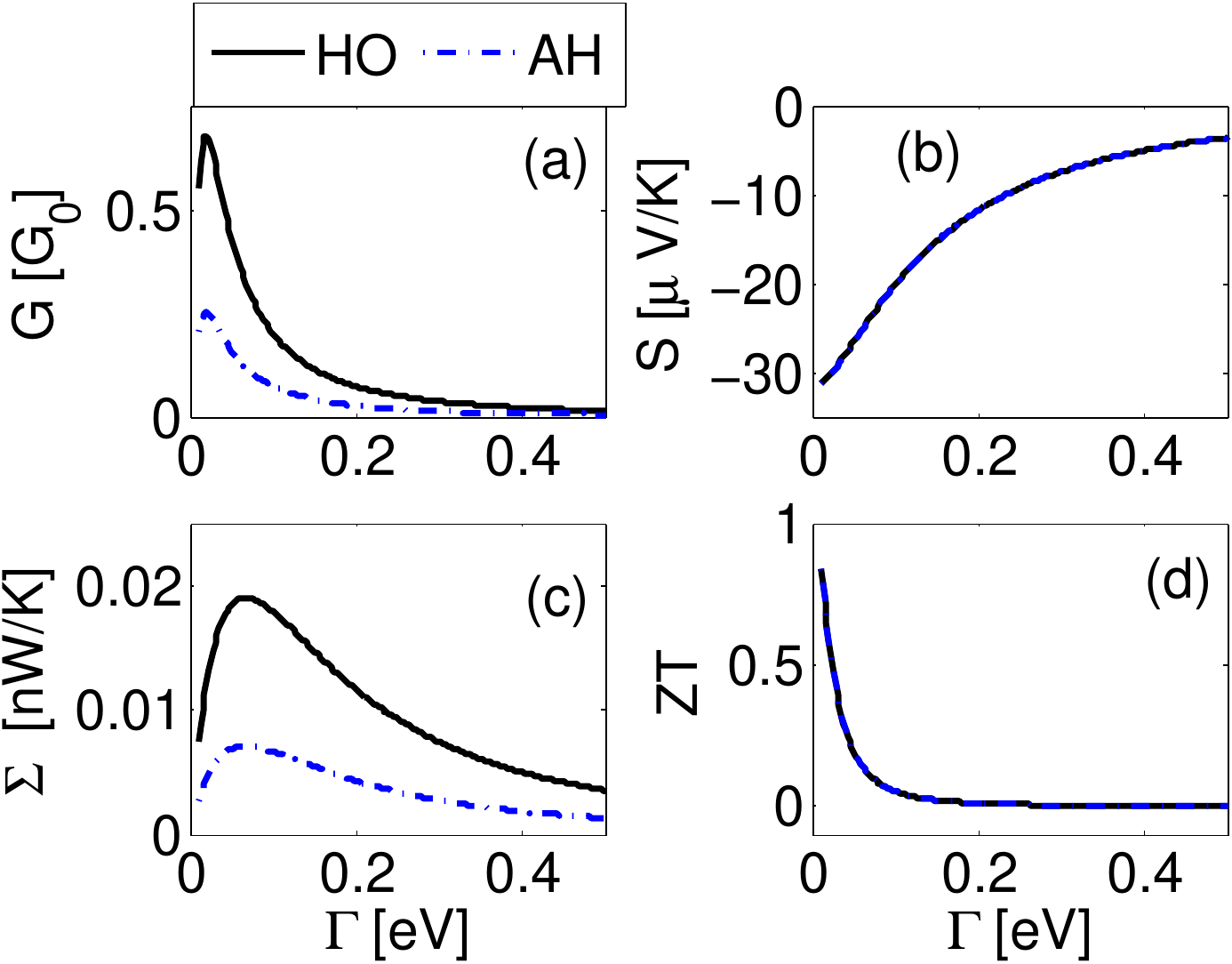}
\end{figure}

\begin{figure}
\caption{Linear response behavior of the donor-acceptor junction as a function of gate voltage.
(a) Electrical conductance, (b) Seebeck coefficient, (c) electronic thermal conductance,
and (d) figure of merit $ZT$.
Parameters are the same as Fig. \ref{fig2} for $\Gamma=0.05$ eV.
}
\label{fig3}
\includegraphics[width=7.5cm]{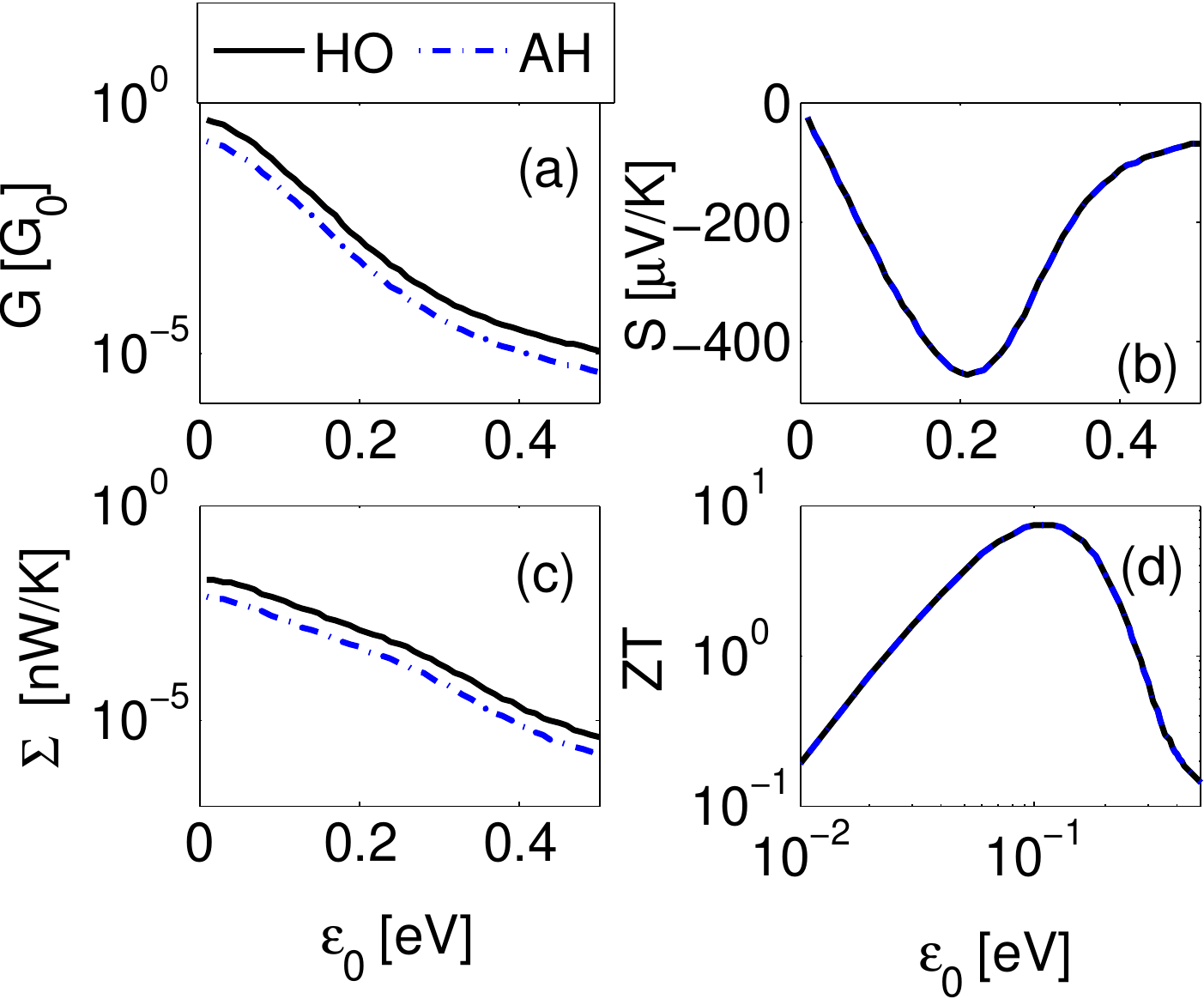}
\end{figure}

\begin{figure}
\caption{Linear response behavior of the donor-acceptor junction as a function of vibrational frequency $\omega_0$
for $\epsilon_0=0.2$ eV, $\Gamma=0.2$ eV,  $g=0.01$ eV, and $T$=300 K.
(a) Electrical conductance, (b) Seebeck coefficient, (c) electronic thermal conductance,
and (d) figure of merit $ZT$.
}
\label{figw0}
\includegraphics[width=7.5cm]{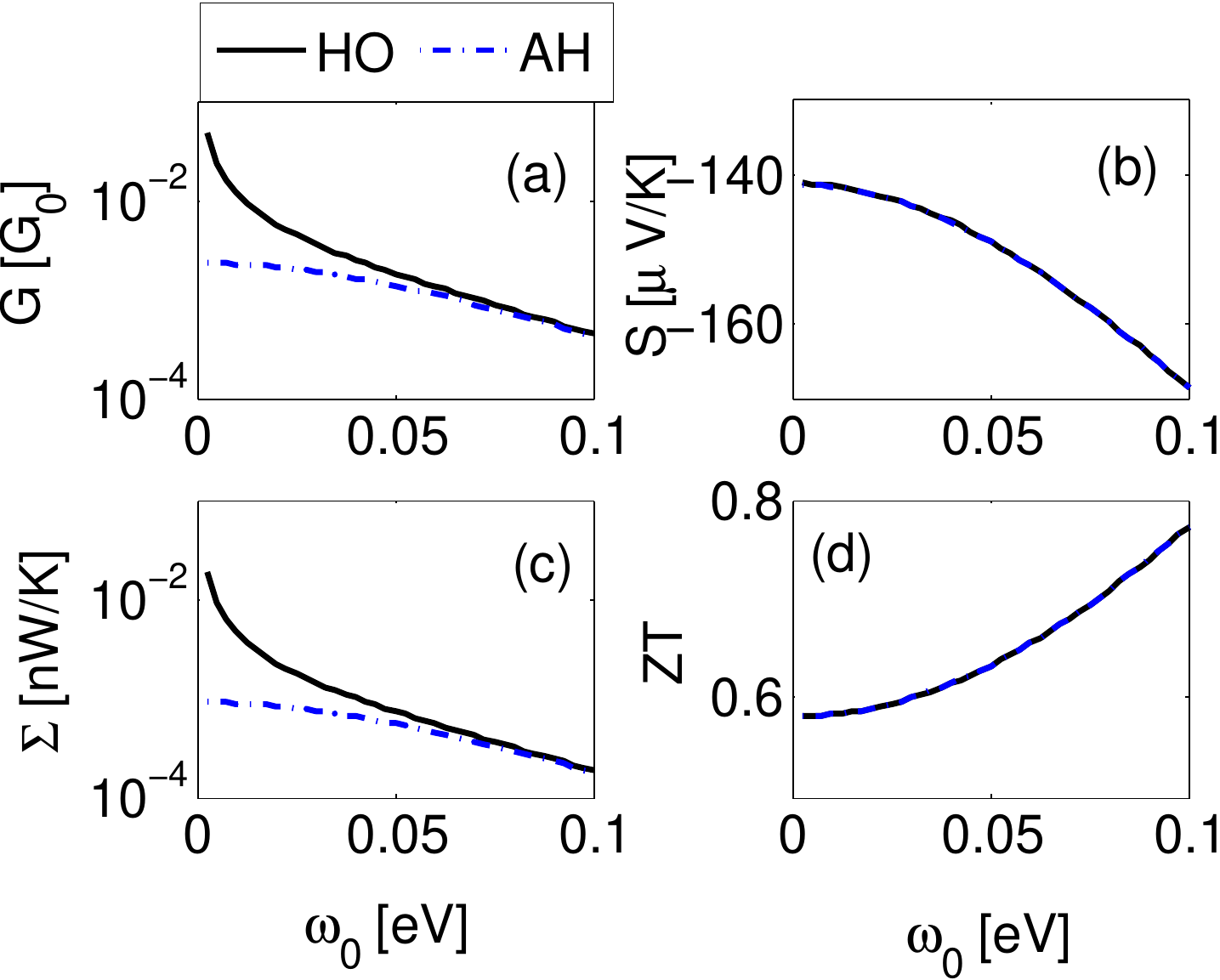}
\end{figure}


Figures  \ref{fig2}-\ref{figw0} display the behavior of $G$, $S$, $\Sigma$ and $ZT$ at room temperature $T=300$ K
for the harmonic and anharmonic-mode junctions. 
In numerical simulations below the phononic contribution to the thermal conductance is
ignored, assuming it to be small compared to its electronic counterpart. A quantitative
analysis of the contribution of the phononic thermal conductance is included in the Discussion Section.
In addition, for simplicity, the junction is made spatially symmetric with
$\Gamma=\Gamma_{L,R}$ and $\epsilon_{0}=\epsilon_{d,a}$.
The currents are given by Eq. (\ref{eq:I}), and we make the following observations:
(i) The harmonic-mode model supports higher currents relative to the two-state case, but  at low temperatures,
$\omega_0/T \gg 1$, when the excitation rate is negligible relative to the relaxation rate,
the two models provide the same results.
(ii) Since the expressions for the currents in the HO and the AH models are proportional to each other,
the resulting thermopower and figure of merit are identical.

Figure  \ref{fig2} displays transport coefficients as a function of metal-molecule hybridization
assuming a resonance situation $k_BT>\epsilon_0$. The conductances show a turnover behavior in accord with
Eq. (\ref{eq:JJ}), growing  with $\Gamma$ for small values $\Gamma<\epsilon_0$,
then falling down approximately as $G,\Sigma \propto \Gamma^{-2}$.
The figure of merit shows a monotonic behavior, increasing
when levels' broadening becomes  small $\Gamma\ll T$
as we approach the so called ``tight coupling" limit in which charge and heat currents
are (optimally) proportional to each other.
%
$ZT$ can be significantly enhanced by tuning the molecule to an off-resonance situation,
$\epsilon_0> k_BT, \Gamma$, see Fig. \ref{fig3}.
We find that the electrical and thermal conductances strongly fall off with $\epsilon_0$, but
the Seebeck coefficient
displays a non-monotonic structure, with a maximum showing up off-resonance \cite{datta},
resulting in a similar enhancement of $ZT$ around $\epsilon_0=0.2$.
It can be proved that the conductances are even functions in gate voltage,
$G(\epsilon_0)=G(-\epsilon_0)$, $\Sigma(\epsilon_0)=\Sigma(-\epsilon_0)$ while $S(\epsilon_0)=-S(-\epsilon_0)$,
resulting in an even symmetry for $ZT$ with gate voltage.

In Fig.  \ref{figw0} we show  transport coefficients as a function of the vibrational frequency.
Parameters correspond to a resonant situation $\epsilon_0/\Gamma=1$. Both $G$ and $\Sigma$ decay exponentially with $\omega_0$
when  $\omega_0>k_BT$. However, the figure of merit only modestly increases with $\omega_0$
in the analyzed range due to the enhancement of $S$ in this region.
The values reported for $ZT$ in figure \ref{figw0} can be
increased
by weakening the metal-molecule coupling energy $\Gamma$.

The maximal efficiency, $\eta_{max}=\eta_C \frac{ \sqrt{ZT+1} -1}{\sqrt{ZT+1}+1}$,
and the efficiency at maximum power, $\eta(P_{max})=\frac{\eta_C}{2}\frac{ZT}{ZT+2}$ \cite{casati13},
are shown in Fig. \ref{figmap} as a function of $\epsilon_0$ and $\Gamma$
for a fixed molecular frequency $\omega_0=0.02$ eV and temperature $T=300$ K.
Here, $\eta_C=1-T_{cold}/T_{hot}$ corresponds to the Carnot efficiency.
By tuning the gate voltage and the molecule-lead hybridization we approach the bounds
$\eta_{max}/\eta_C\rightarrow 1$,  $\eta(P_{max})/\eta_C\rightarrow 1/2$ \cite{casati13}.
Particularly, for $\Gamma\sim 0.01$ eV we receive $\eta_{max}/\eta_C=0.8$ at
the energy $\epsilon_0=0.15$.


\begin{figure}
\caption{Contour plot of linear response efficiencies as a function of hybridization $\Gamma$ and
electronic energies (or gate voltage) $\epsilon_0$.
(a) Maximum efficiency. (b) Efficiency at maximum power.
Parameters are $\omega_0=0.02$ eV and  $T$= 300 K.
}
\label{figmap}
\hspace{5mm}
\includegraphics[width=10cm]{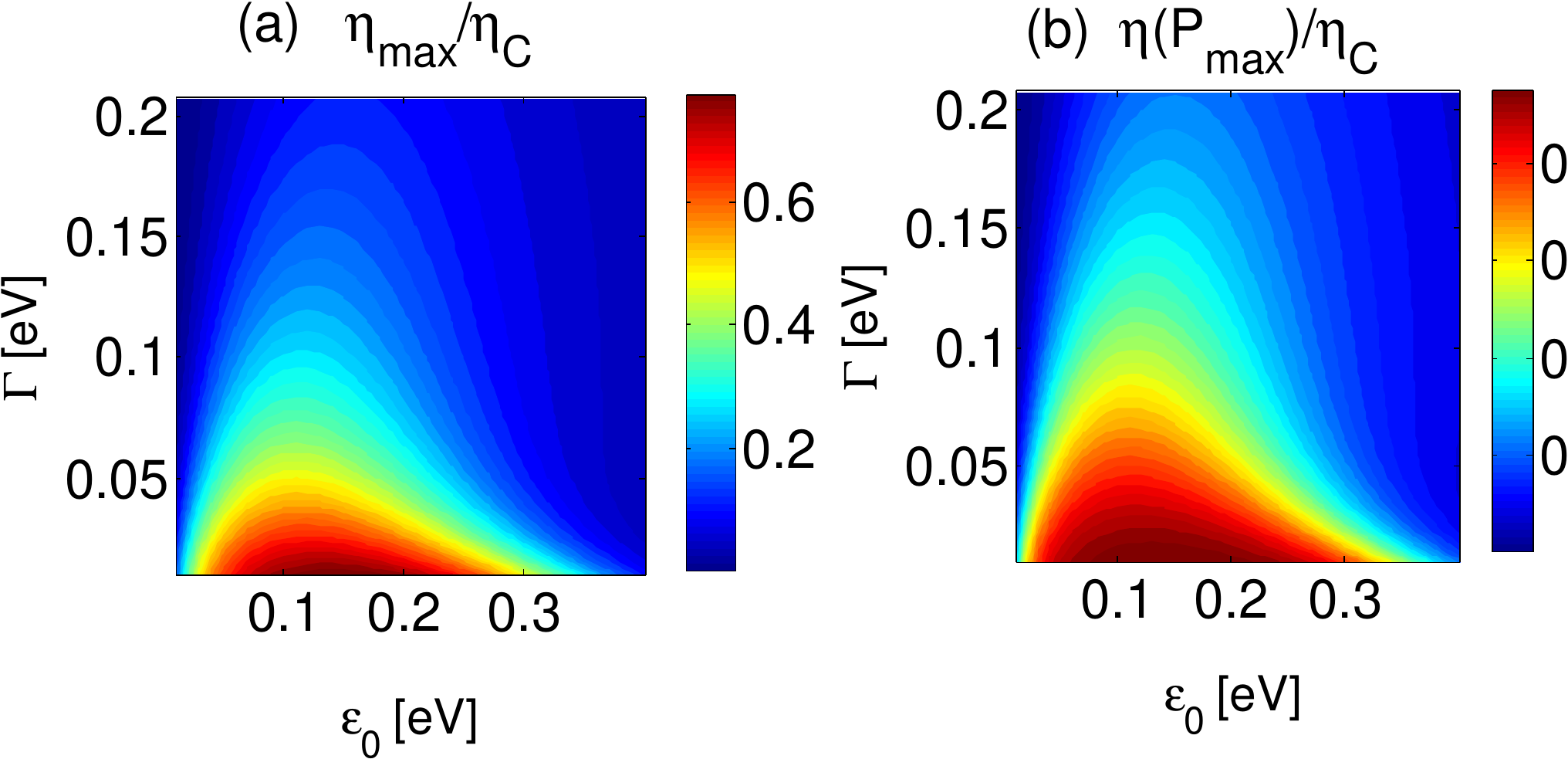}
\end{figure}

\subsection{Nonlinear performance}

Nonlinear thermoelectric phenomena  are anticipated to enhance thermoelectric response \cite{shakouri4}.
Elastic scattering theories of nonlinear thermoelectric transport have been developed e.g.
in Refs.  \cite{nonlinear,sanchez,Linke,Whitney},
accounting for many-body effects in a phenomenological manner.
Only few studies had considered this problem
with explicit electron-phonon interactions,  based on the Anderson-Holstein model \cite{wegewijs}
or Fermi-Golden rule expressions \cite{JianHua2}.

In Fig. \ref{figNL} we simulate the current-voltage characteristics
and the resulting efficiency of the D-A junction
beyond linear response, by directly applying Eq. (\ref{eq:I}).
As discussed in previous investigations \cite{Lu,SegalSF1,SegalSF2},
the molecular junction may break down far from equilibrium due to the development of
``vibrational instability".
This over-heating effect occurs when (electron-induced) vibrational excitation rates
exceed relaxation rates. To cure this physical problem, we  allow the particular vibrational mode of frequency $\omega_0$
to relax its excess energy
to a secondary phonon bath of temperature $T_{ph}$. This can be done rigorously
at the level of the quantum master equations and within the NEGF technique \cite{SegalSF1,BijayCGF} to yield the rates
$k_{d}= k_{d}^{L \to R} + k_{d}^{R \to L}+\Gamma_{ph}(\omega_0)[n_{ph}(\omega_0)+1]$,
$k_{u}= k_{u}^{L \to R} + k_{u}^{R \to L}+\Gamma_{ph}(\omega_0) n_{ph}(\omega_0)$,
with $n_{ph}(\omega_0)=[e^{\omega_0/k_BT_{ph}}-1]^{-1}$
and a damping term $\Gamma_{ph}(\omega_0)$.
%
Interestingly, we confirmed (not shown)
that this additional energy relaxation process does not modify the thermoelectric efficiency displayed in Fig. \ref{figNL}(b).

\begin{figure}
\vspace{3mm}
\caption{Transport beyond linear response. (a)
current voltage characteristics for the harmonic (full) and anharmonic (dashed) mode models.
(b) Heat to work conversion efficiency [Eq. (\ref{eq:eta})].
Parameters are $\omega_0=0.02$, $\epsilon_0=0.2$, $g$=0.01, $\Gamma=0.1$, $\Gamma_{ph}=0.002$ in units of  eV, and
$T_L$= 300 K, $T_R$= 800 K and $T_{ph}$=300 K.
}
\label{figNL}
\includegraphics[width=7cm]{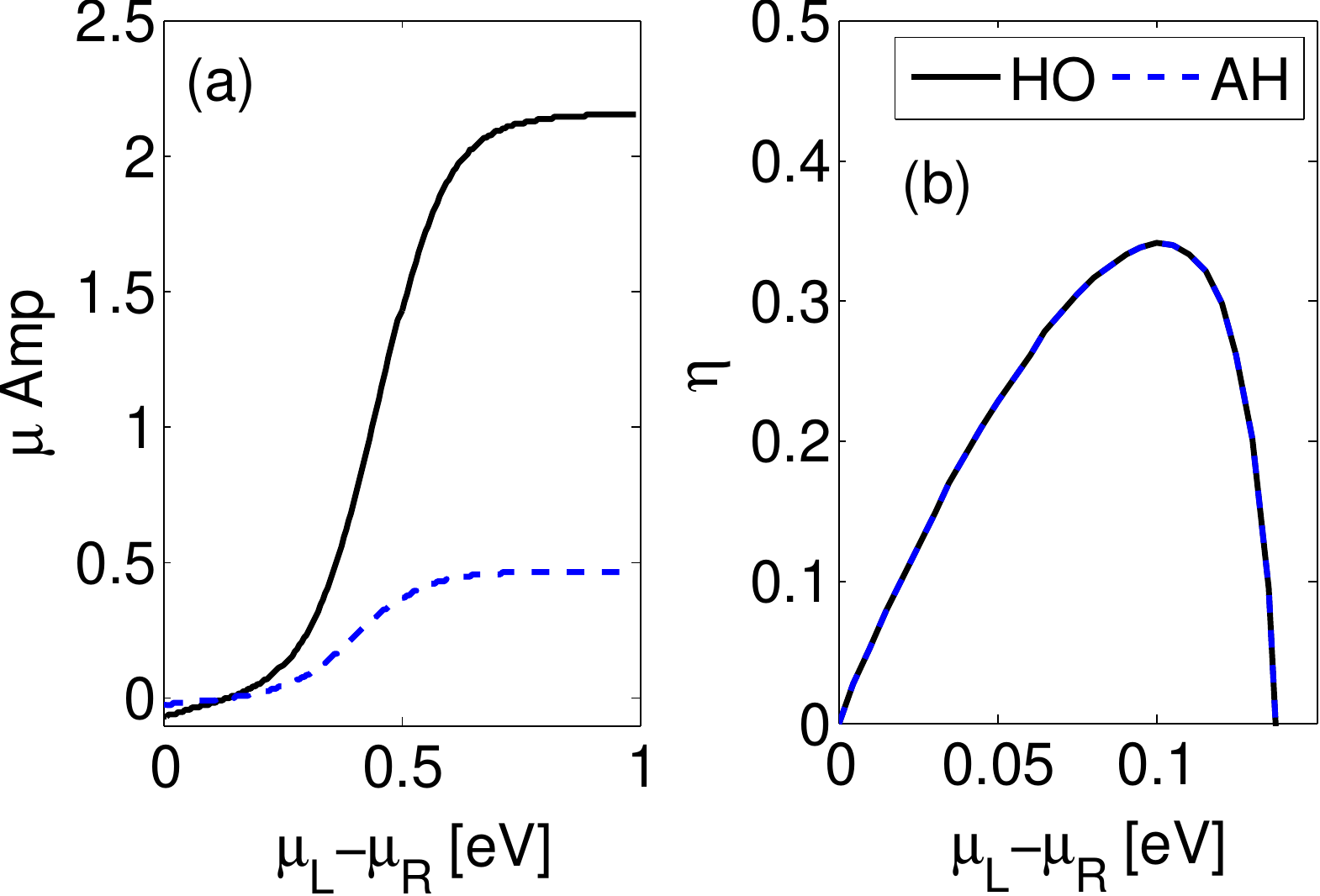}
\end{figure}

\section{Discussion and Prospect}

We focused on two-site electronic junctions in which electron transfer between sites is assisted by a particular
mode, harmonic or anharmonic (two-state system).
The complete information over steady state transport behavior
is catered by the cumulant generating function, which we provide here for the HO and the AH mode models,
valid under the approximation of weak electron-vibration interaction.
We explored linear-response properties, the electrical and thermal conductances $G$ and $\Sigma$,
 as well as the Seebeck coefficient $S$, the thermoelectric figure of merit $ZT$,
and the resulting efficiency. We further
examined current-voltage behavior and the heat-to-work conversion efficiency far-from-equilibrium. We
found that $G$ and $\Sigma$ (more generally, the charge and energy currents)
are sensitive to the properties of the mode,
while $S$ and $ZT$ are insensitive to whether we work with a harmonic mode or a truncated two-state model.
Several comments are now in place:

(i) {\it Genuine anharmonicity.} We examined the role of mode
anharmonicity by devising a two-state impurity model.
It should be emphasized that in the context of molecules,
the two-state impurity does not well represent vibrational
anharmonicity at high temperatures, as many states should then contribute.
Furthermore, it misses an explicit parameter tuning the potential anharmonicity.
However, the AH model allows for a first indication on how
deviations from harmonicity reflect in transport behavior.
The HO and AH models have similar CGFs, yielding currents
which are proportional, thus an identical thermoelectric efficiency.
It can be readily shown that an $n$-state truncated HO provides a figure of merit identical
to the infinite-level HO model, but
it is interesting to perform more realistic calculations and consider
e.g. a morse potential to represent a physical anharmonic
 molecular vibration. In this case, an analytical form for the
CGF is missing, but one could still derive the charge current directly from a quantum master equation formalism,
to obtain the performance of the system.
We expect that with a genuine anharmonic potential,
$S$ and $ZT$ would show deviations from the harmonic limit,
as different pathways for transport open up.
Overall, we believe that our results here indicate on the minor role
played by mode anharmonicity in determining heat-to-work conversion efficiency.

(ii) {\it Direct tunneling.}
Our analysis was performed while neglecting direct electron tunneling between the D and A sites. This effect could be
approximately re-instituted
by assuming that coherent transport proceeds in parallel to phonon-assisted conduction,
accounting for the coherent contribution using a Landauer expression, see e.g., Refs. \cite{Markussen,JianHua2}.
Indeed, path integral simulations indicted that in the D-A model, coherent and the incoherent
contributions are approximately additive \cite{SegalSF2}.

(iii) {\it Strong electron-phonon interaction.}
The CGFs (\ref{eq:CGFHO})-(\ref{eq:CGFAH}) are exact to all orders in the metal-molecule hybridization but
perturbative (to the lowest nontrivial order) in the electron phonon coupling $g$.
This is evident from the structure of the rate constants in Eq. (\ref{eq:rates}),
as electron transfer is facilitated by the absorption/emission
of a single quantum $\omega_0$.
In numerical simulations we typically employed $g=0.01$ eV and $\omega_0=0.02$ eV.
This value for $g$ may seem large given the perturbative nature of our treatment
requiring ${g}/{\omega_0}\ll1$. However, since in the present weak-coupling limit
the current simply scales as $g^2$, Eq. (\ref{eq:I}), our simulations in this work
are representative, and can be immediately translated for other values for $g$.
It is of interest to generalize our results and study the performance of the junction
with strong electron-phonon interaction, e.g. by using a polaronic transformation
\cite{JensK,Hartle,strongKomnik,strongEsp,strongLei}.

(iv) {\it Phononic thermal conductance.}
We studied here electron transfer through molecular junctions, but did not discuss phonon transport
characteristics across the junction, mediated by molecular vibrational modes.
Consideration of the phononic thermal conductance $\kappa_{ph}$  is particularly important for a
reliable estimate of $ZT$,
as the thermal conductance $\Sigma$ should include contributions from both electrons and phonons.
We now estimate $\kappa_{ph}$.
The quantum of thermal conductance, an upper bound for ballistic conduction, is given by
$\kappa_Q\equiv \pi k_B^2T/6\hbar$ \cite{Rego98}. At room temperature, this provides $\kappa_Q=0.28$ nW/K,
which exceeds the
electronic thermal conductance obtained in our simulations,
to dominate the total thermal conductance and predict (significantly) lower values for the figure of merit.
However, one should recognize that at high temperatures the ballistic bound for phonon thermal conductance
is far from being saturated
as was recently demonstrated in Ref. \cite{Ed}. In particular,
the phononic thermal conductance of a two-level junction
was evaluated exactly in Ref. \cite{Saito13}, and it significantly falls below the harmonic bound \cite{Ed}.

For a concrete estimate, we adopt the perturbative
(weak mode-thermal bath) expression  for the
phononic current through a two-state junction
developed in Ref. \cite{SegalR05} and further examined in Ref. \cite{Boudjada},
\bea
j_{ph}^{AH}=g_{ph}\omega_0\frac{[n_R(\omega_0)-n_L(\omega_0)]}{ [2+2n_R(\omega_0)+ 2n_L(\omega_0)]}.
\label{eq:kph}
\eea
Here,
$g_{ph}$ is the interaction strength of the local vibrational mode to the phononic environments at the two terminals
(assuming identical interaction strengths).
The baths are  characterized by their  Bose-Einstein distribution functions $n_{\nu}(\omega)$.
In the case of a local harmonic mode, Eq. (\ref{eq:kph}) holds, only missing its denominator.
Using $\omega_0=0.02$ eV,
we receive an estimate for the phononic thermal conductance $\kappa_{ph}^{AH}\equiv j_{ph}^{AH}/\Delta T$,
$\kappa_{ph}^{AH}\sim 3.5 \times g_{ph}$ nW/K. Thus, as long as the mode-bath coupling $g_{ph}$ is taken as weak,
for example, $g_{ph}<5$ meV for the data of Fig. \ref{fig2},
the electronic contribution to the thermal conductance
dominates the total thermal conductance and our simulations are intact.
Similar considerations hold for harmonic-mode junctions.
Proposals to reduce the coherent phononic thermal conductance by quantum interference effects \cite{MarkussenP} and
through-space designs \cite{gemmap} could be further considered.

(v) {\it High order cumulants.} The cumulant generating functions, Eqs. (\ref{eq:CGFHO}), and (\ref{eq:CGFAH}),
contain significant information. For example, one could
examine the (zero-frequency) current noise, to find out to  what extent it can reveal microscopic molecular information.

(vi) {\it Methodology development.}
The cumulant generating function of the HO model was derived from an NEGF approach \cite{BijayCGF}.
The corresponding function for the AH model was reached from a master equation calculation \cite{SegalSF1,BijayCGF}.
Both treatments are perturbative to second order in the electron-vibration interaction; we take into account
all electron scattering processes that are facilitated by
the absorption or emission of a {\it single} quantum $\omega_0$.
It is yet surprising to note on the
direct correspondence between NEGF and master equation results, as derivations proceeded on completely
different lines. In particular,
the NEGF approach was done at the level of the RPA approximation to guarantee the validity of the fluctuation theorem.
The master equation approach has been employed before to study currents (first cumulants) in the HO model
\cite{SegalSF1},
showing exact agreement with NEGF expressions presented here.
This agreement, as well as supporting path integral simulations \cite{SegalSF2},
indicate on the accuracy and consistency of the master equation in the present model.
Given its simplicity and transparency,
it is of interest to extend this method and examine
higher order processes in perturbation theory, to gain further insight on the role of electron-vibration
interaction in molecular conduction.

(vii) {\it Efficiency fluctuations.} We focused here only on averaged-macroscopic quantities.
However,  in small systems fluctuations in input heat and output power are significant, resulting in
``second law violations" as predicted from the fluctuation theorem  \cite{fluctR},
to e.g. grant efficiencies exceeding the
thermodynamic bound. To analyze the distribution of efficiency,
the concept of ``stochastic efficiency" has been recently coined and examined \cite{me1,me2}.
In a separate contribution \cite{BijayCGF} we extend the present analysis and describe the characteristics of
the stochastic efficiency in our model, particularly, we explore signatures of mode anharmonicity
in the statistics of efficiency.

(viii) {\it Molecular calculations.} It is of interest to employ our expressions and examine heat-to-work conversion
efficiency in realistic molecular junctions. Our results demonstrate that the conversion efficiency can be improved by
working in the off-resonant limit, $\Gamma/\epsilon_0\ll1$, as well, when tuning $\epsilon_0$ through a gate voltage to
$\epsilon_0/k_BT\sim 5$. In such situations, the figure of merit $ZT$
can be made large since one can tune $S$ to large values (though the conductances are small).
This suggests that in the D-A class of molecules one should focus on enhancing the thermopower
as a promising mean for making an overall improvement in efficiency.

In our ongoing work we are pursuing some of these topics.
The derivation of the CGFs employed here
and the behavior of efficiency fluctuations in linear response, and beyond that, are detailed in Ref. \cite{BijayCGF}.
In Ref. \cite{JianHua2} we derive thermoelectric transport coefficients for the dissipative D-A model,
beyond linear response,
and describe the operation of thermoelectric diodes and transistors.

\begin{acknowledgements}
The work of DS and BKA was supported by an NSERC
Discovery Grant, the Canada Research Chair program,
and the CQIQC at the University of Toronto.
JHJ acknowledges support from the faculty start-up funding of Soochow
University.
\end{acknowledgements}


\end{document}